\documentclass[
    ,final            % use final for the camera ready runs
  ]
  {aipproc}
\usepackage{bm}
\layoutstyle{6x9}

\newcommand{\eq}{{\,=\,}}

\newcommand{\half}{\textstyle{\frac{1}{2}}}

\begin{document}

\title{Thermalization at RHIC}

\author{Ulrich Heinz}{
  address={Physics Department, The Ohio State University, 174 West 18th 
           Avenue, Columbus, OH 43210, USA}
}

\begin{abstract}
Ideal hydroynamics provides an excellent description of all aspects of 
the single-particle spectra of all hadrons with transverse momenta 
below about 1.5-2 GeV/$c$ at RHIC. This is shown to require rapid 
local thermalization at a time scale below 1 fm/$c$ and at energy 
densities which exceed the critical value for color deconfinement by 
an order of magnitude. The only known thermalized state at such 
energy densities is the quark-gluon plasma (QGP). The rapid 
thermalization indicates that the QGP is a strongly interacting 
liquid rather than the weakly interacting gas of quarks and gluons 
that was previously expected.
\end{abstract}

\maketitle

%%%%%%%%%%%%%%%%%%%%%%%%%%%%%%%%%%%%%%%%%%%%
%% MAINMATTER
%%%%%%%%%%%%%%%%%%%%%%%%%%%%%%%%%%%%%%%%%%%%

%%%%%%%%%%%%%%%%%%%%%%%%%%%%%%%%%%%%%%%%%%%%%%%%%%%%%%%%%%%%%%%%%%%%%%%%%%%
\subsection{Collective flow as an indicator of thermalization}
%%%%%%%%%%%%%%%%%%%%%%%%%%%%%%%%%%%%%%%%%%%%%%%%%%%%%%%%%%%%%%%%%%%%%%%%%%%

Transverse flow in heavy-ion collisions is an unavoidable consequence 
of thermalization. Thermalization generates thermodynamic pressure in 
the matter created in the collision, which acts against the surrounding
vacuum and causes rapid collective expansion (``flow'') of the reaction 
zone. Since the quark-gluon plasma (QGP) is a thermalized system of 
deconfined quarks, antiquarks, and gluons, collective flow is a 
necessary result of QGP formation in heavy-ion collisions, and its 
absence could be taken as proof that no such plasma was ever formed.
Its presence, however, does not automatically signal QGP formation.
Detailed studies of the observed final state flow pattern are necessary 
to convince oneself that the reflected time-integrated pressure history 
of the collision region indeed requires a thermalized state in the 
early collision stage whose pressure and energy density are so high 
that it can no longer be mistaken as consisting of conventional 
hadronic matter. 

Whereas {\em radial flow} (the azimuthally symmetric component of collective 
expansion transverse to the beam direction) integrates over the entire 
pressure history of the expanding fireball, anisotropic {\em elliptic 
flow} \cite{O92} (quantified by the second harmonic coefficient 
$v_2(y,p_\perp;b)$ of a Fourier expansion in $\phi_p$ of the measured 
hadron spectrum $dN/(dy\,p_\perp dp_\perp\,d\phi_p)$ \cite{VZ96}) is 
strongly weighted towards the very early stages of the expansion
\cite{S97}. The higher the initial energy density, the less contributions 
it receives from the late, hadronic stage of the collision. At RHIC 
energies, elliptic flow almost saturates \cite{KSH00} before the energy 
density has dropped to the critical value $e_{\rm cr} \approx 1$\,GeV/fm$^3$
($T_{\rm cr}\approx 170$\,MeV) where normal hadrons can begin to form 
\cite{Karsch_QM01}. At higher LHC energies, the elliptic flow is expected
to peak even before the onset of hadronization (see curve $c$ in Fig. 7 of 
\cite{KSH00}). At sufficiently high collision energies, elliptic flow is
thus a $\bm{QGP\ signature}$, probing the QGP equation of state $p(e)$. Its
advantage over other ``early signatures'' is that it affects the bulk of 
the hadrons and can thus be measured differentially with high statistical 
accuracy.

The reason why elliptic flow must develop {\em early} in the collision
is easy to understand. Since individual nucleon-nucleon collisions 
produce azimuthally symmetric momentum spectra, any final state momentum 
anisotropies must be generated dynamically during the nuclear reaction. 
They require the existence of an initial {\em spatial} anisotropy of the 
reaction zone, either by colliding deformed nuclei such as U+U
\cite{KSH00,Sh00,Li00}, or by colliding spherical nuclei at non-zero 
impact parameter $b{\,\ne\,}0$ (the practical method of choice so far).   
Final state interactions within the produced matter transfer the
initial spatial anisotropy onto a final momentum anisotropy.
Microscopic transport calculations 
\cite{Zhang:1999rs,Molnar:2001ux} show a monotonic dependence of $v_2$ 
on the opacity (density times scattering cross section) of the produced 
matter which is inversely related to its thermalization time. These studies
strongly suggest that, for a given initial spatial anisotropy $\epsilon_x$, 
the maximum momentum-space response $v_2$ is obtained in the {\em ideal 
hydrodynamic limit} which assumes perfect local thermal equilibrium at 
every space-time point (i.e. a thermalization time which is much shorter 
than any macroscopic time scale in the system). Any significant delay of 
thermalization (modelled, for example, as an initial free-streaming stage) 
causes a decrease of the initial spatial anisotropy without concurrent 
build-up of momentum anisotropies, thereby reducing the finally observed 
elliptic flow signal \cite{KSH00}.

In this talk I will present results from hydrodynamic simulations of
hadronic spectra and elliptic flow at RHIC energies. We will see that
the hydrodynamic approach provides an excellent quantitative description 
of the bulk of the data and fails only for very peripheral Au+Au collisions
and/or at high $p_\perp{>}1.5{-}2$\,GeV/$c$. That the hydrodynamic approach
fails if the initial nuclear overlap region becomes too small or the 
transverse momentum of the measured hadrons becomes too large is not 
unexpected. However, where exactly hydrodynamics begins to break down 
gives important information about the microscopic rescattering dynamics. 
What is really surprising is that the hydrodynamic approach works so well 
in central and semi-central collisions where it quantitatively reproduces
the momenta of more than 99\% of the particles. Below $p_\perp\eq1.5$\,GeV/$c$
the elliptic flow data \cite{Ackermann:2001tr,PHENIXv2,PHOBOSv2} 
actually exhaust the hydrodynamically predicted 
\cite{KSH00,Teaney:2001cw,Kolb:2001fh,Huovinen:2001cy,KHHET} 
upper limit. The significance of this agreement can hardly be overstressed,
since it implies one of the biggest surprises so far at RHIC: the produced
matter (which I call QGP since this is only known viable concept of 
thermalized matter at $e\sim (10{-}20)\,e_{\rm cr}$) is not the originally
expected weakly interacting gas of quarks and gluons ($\bm{wQGP}$ for 
``weakly interacting quark-gluon plasma''), but a strongly coupled liquid 
with extremely small viscosity ($\bm{sQGP}$ for ``strongly interacting 
quark-gluon plasma'' \cite{TDLee}). In fact, upper limits \cite{T03} on 
the dimensionless ratio of viscosity to entropy density,
$\eta/s$, based on an analysis of the RHIC elliptic flow data, 
indicate that the sQGP is less viscous, by about an order of magnitude, 
than even liquid helium below the transition to superfluidity. {\it The 
sQGP is therefore the most ideal fluid ever observed}!   

%%%%%%%%%%%%%%%%%%%%%%%%%%%%%%%%%%%%%%%%%%%%%%%%%%%%%%%%%%%%%%%%%%%%%%%%%%%
\subsection{Hydrodynamic expansion in heavy-ion collisions}
%%%%%%%%%%%%%%%%%%%%%%%%%%%%%%%%%%%%%%%%%%%%%%%%%%%%%%%%%%%%%%%%%%%%%%%%%%%

The natural language for describing collective flow phenomena is 
hydrodynamics. Its equations control the space-time evolution of
the pressure, energy and particle densities and of the local fluid 
velocity. The system of hydrodynamic equations is closed by specifying
an {\em equation of state} which gives the pressure as a function of 
the energy and particle densities. In the ideal fluid (non-viscous) 
limit, the approach assumes that the microscopic momentum distribution 
is thermal at every point in space and time (note that this does not 
require chemical equilibrium -- chemically non-equilibrated situations 
can be treated by introducing into the equation of state non-equilibrium
chemical potentials for each particle species 
\cite{Rapp02,Hirano02,Teaney02}). Small deviations from local thermal 
equilibrium can in principle be dealt with by including viscosity, 
heat conduction and diffusion effects, but such a program is made 
difficult in practice by a number of technical and conceptual 
questions \cite{Rischke:1998fq} and has so far not been successfully
applied to relativistic fluids. Stronger 
deviations from local thermal equilibrium require a microscopic 
phase-space approach (kinetic transport theory), but in this case the 
concepts of equation of state and local fluid velocity field 
themselves become ambiguous, and the direct connection between flow 
observables and the equation of state of the expanding matter is lost. 

The assumption of local thermal equilibrium in hydrodynamics is an 
external input, and hydrodynamics offers no direct insights about the 
equilibration mechanisms. It is clearly invalid during the initial 
particle production and early recattering stage, and it again breaks 
down towards the end when the matter has become so dilute that 
rescattering ceases and the hadrons ``freeze out''. The hydrodynamic 
approach thus requires a set of {\em initial conditions} for the 
dynamic variables at the earliest time at which the assumption of
local thermal equilibrium is applicable, and a {\em ``freeze-out 
prescription''} at the end. For the latter we use the Cooper-Frye algorithm 
\cite{Cooper:1974mv} which implements an idealized sudden transition 
from perfect local thermal equilibrium to free-streaming. This is not 
unreasonable because freeze-out (of particle species $i$) is controlled 
by a competition between the local expansion rate $\partial\cdot u(x)$ 
(where $u^\mu(x)$ is the fluid velocity field) and the local scattering 
rate $\sum_j \langle \sigma_{ij} v_{ij}\rangle \rho_j(x)$ (where the sum
goes over all particle species with densities $\rho_j(x)$ and 
$\langle \sigma_{ij} v_{ij}\rangle$ is the momentum-averaged transport 
cross section for scattering between particle species $i$ and $j$, 
weighted with their relative velocity); while the local expansion rate 
turns out to have a rather weak time-dependence, the scattering
rate drops very steeply as a function of time, due to the rapid
dilution of the particle densities $\rho_j$ \cite{Schnedermann:1994gc},
causing a rapid transition to free-streaming. -- A better algorithm 
\cite{Bass:2000ib,Teaney:2001cw} switches from a hydrodynamic description 
to a microscopic hadron cascade at or shortly after the quark-hadron 
transition, before the matter becomes too dilute, and lets the cascade 
handle the freeze-out kinetics. This also correctly reproduces the 
final chemical composition of the fireball, since the particle 
abundances already freeze out at hadronization, due to a lack of 
particle-number changing inelastic rescattering processes in the 
hadronic phase. The resulting radial flow patterns 
\cite{Teaney:2001cw} from such an improved freeze-out algorithm 
don't differ much from our simpler Cooper-Frye based approach. 

%%%%%%%%%%%%%%%%%%%%%%%%%%%%%%%%%%%%%%%%%%%%%%%%%%%%%%%%%%%%%%%%%%%%%%%%%%%
\subsection{Hydrodynamic radial flow and RHIC particle spectra}
%%%%%%%%%%%%%%%%%%%%%%%%%%%%%%%%%%%%%%%%%%%%%%%%%%%%%%%%%%%%%%%%%%%%%%%%%%%

We have solved the relativistic equations for ideal hydrodynamics, as 
described in \cite{KSH00}. To simplify the numerical task, we imposed 
boost-invariant longitudinal expansion analytically \cite{O92,Bj83}. 
As long as we focus on the transverse ex\-pan\-sion dynamics near 
midrapidity (the region which most RHIC experiments cover best), 
this does not give up any essential physics.

We use an EOS which is constructed by matching
a free ideal quark-gluon gas above $T_{\rm cr}$ to a realistic hadron 
resonance gas \cite{LRH88b,SHKRPV97} below $T_{\rm cr}$, using a Maxwell 
construction and fixing $T_{\rm cr}\eq165$\,MeV to reproduce lattice QCD 
results \cite{Karsch_QM01}. The Maxwell construction leads to an 
artificial first order transition, with a latent heat of 1.15\,GeV/fm$^3$,
whereas the lattice QCD data indicate either a very weak first order
transition or a rapid, but smooth, crossover. As long as this crossover
is very steep (as the lattice data indicate \cite{Karsch_QM01}), the
dynamical consequences of our first-order idealization of the EOS 
are not expected to be significant.

We have used two variants of the hadron resonance gas below $T_{\rm cr}$:
In our first sets of calculations 1999-2002, we assumed the hadron 
resonance gas to be in chemical equilibrium until kinetic freeze-out. 
RHIC data on particle abundance ratios show, however, that hadronic 
particle yields freeze out directly at $T_{\rm cr}$ \cite{BMMRS01}, due 
to inefficiency of particle-changing inelastic reactions in the 
relatively dilute and rapidly expanding hadron gas phase below 
$T_{\rm cr}$. More recent hydrodynamic calculations \cite{Hirano02,KR03}
therefore include non-equilibrium chemical potentials 
\cite{Rapp02,Hirano02,Teaney02} ensuring number conservation for
individual (stable) hadron species below $T_{\rm cr}$. This does not
affect the equation of state $p(e)$ and therefore leads to the same
flow pattern as the chemical equilibrium EOS \cite{Hirano02}. What does
change, however, is the relationship between energy density $e$ and
temperature $T$ since in the chemical non-equilibrium case more of
the energy is stored in the rest masses of heavy baryons and antibaryons 
(which are not allowed to annihilate as the temperature drops). The same
decoupling energy density $e_{\rm dec}\eq0.075$\,GeV/fm$^3$ obtained
from a fit to RHIC data (see next paragraph) therefore corresponds to
$T_{\rm dec}\eq130$\,MeV in the chemical equilibrium case, but drops to
the much lower value $T_{\rm dec}\eq100$\,MeV if hadronic {\em chemical} 
freeze-out at $T_{\rm cr}\eq165$\,MeV is properly taken into account 
\cite{Hirano02,KR03}.

The initial and final conditions for the hydrodynamic evolution are fixed
by fitting the total charged multiplicity $dN_{\rm ch}/dy$ and the pion 
and proton spectra at midrapidity in {\em central} ($b\eq0$) collisions.
(All calculated hadron spectra include feeddown from decays of unstable 
hadron resonances \cite{SKH90}.) Since these two hadrons have quite 
different masses, their spectra allow for the independent extraction of 
the temperature and average radial flow velocity at freeze-out
\cite{Lee:1990sk}. The 
freeze-out temperature together with the total charged multiplicity 
determines the total fireball entropy which in ideal hydrodynamics is 
preserved during the expansion. During the early, predominantly 
longitudinal expansion stage, where the fireball volume increases 
linearly with proper time $\tau$, this constraint fixes the product
$\tau\cdot\int d^2r_\perp\,s(\bm{r}_\perp,\tau)$ where $s(\bm{r}_\perp,\tau)$
is the entropy density distribution in the plane transverse to the beam.
Since the initial shape of this distribution is fixed by the nuclear 
overlap geometry, using a Glauber model ansatz \cite{KHHET}, this 
constraint determines the normalization of $s(\bm{r}_\perp,\tau_{\rm eq})$
(i.e. the central entropy density $s_0{\eq}s(\bm{0},\tau_{\rm eq})$)
as a function of the initial thermalization time $\tau_{\rm eq}$. The
Glauber model contains one additional parameter: the ratio of ``soft'' and 
``hard'' components of initial particle production. Since these scale with 
the transverse density of wounded nucleons and binary nucleon-nucleon 
collisions, respectively, their ratio can be determined from the 
collision centrality dependence of the produced charged multiplicity 
$dN_{\rm ch}/dy$ \cite{KHHET}. We use 25\% hard and 75\% soft 
contributions to the initial entropy production \cite{Heinz:2001xi}. 
The initial thermalization time $\tau_{\rm eq}$, finally, is fixed by 
the need for the hydrodynamic evolution to have enough time to generate
the finally observed radial flow. At RHIC energies, we find that this 
requires thermalization times $\tau_{\rm eq}{\,\leq\,}0.6$\,fm/$c$.
In our calculations we take the upper limit of this interval. In Au+Au 
collisions at $\sqrt{s}\eq100\,A$\,GeV, the initial central entropy 
density at this time is $s_0\eq110$\,fm$^{-3}$, corresponding to an 
initial central energy density\footnote{When averaged over the 
transverse profile, this corresponds to 
$\langle e\rangle\eq13$\,GeV/fm$^3$ which is still a good order of 
magnitude above the critical value for deconfinement, 
$e_{\rm cr}\eq0.6-1$\,GeV/fm$^3$ \cite{Karsch_QM01}.} 
$e_0\eq30$\,GeV/fm$^3{\,\approx\,}30\,e_{\rm cr}$ and an initial central 
temperature $T_0\eq360\,{\rm MeV}{\,\approx\,}2\,T_{\rm cr}$.\footnote{If 
the initial QGP is not chemically equilibrated, but rather dominated by 
gluons, this initial temperature could be as high as 460\,MeV.} 

%%%%%%%%%%%%%%%%%%%%%%%%% Fig. 1 %%%%%%%%%%%%%%%%%%%%%%%%%%%%%%%%%%%%%%%%%%%%
\begin{figure}[htb]
\begin{minipage}[t]{70mm}
 \includegraphics[width=70mm]{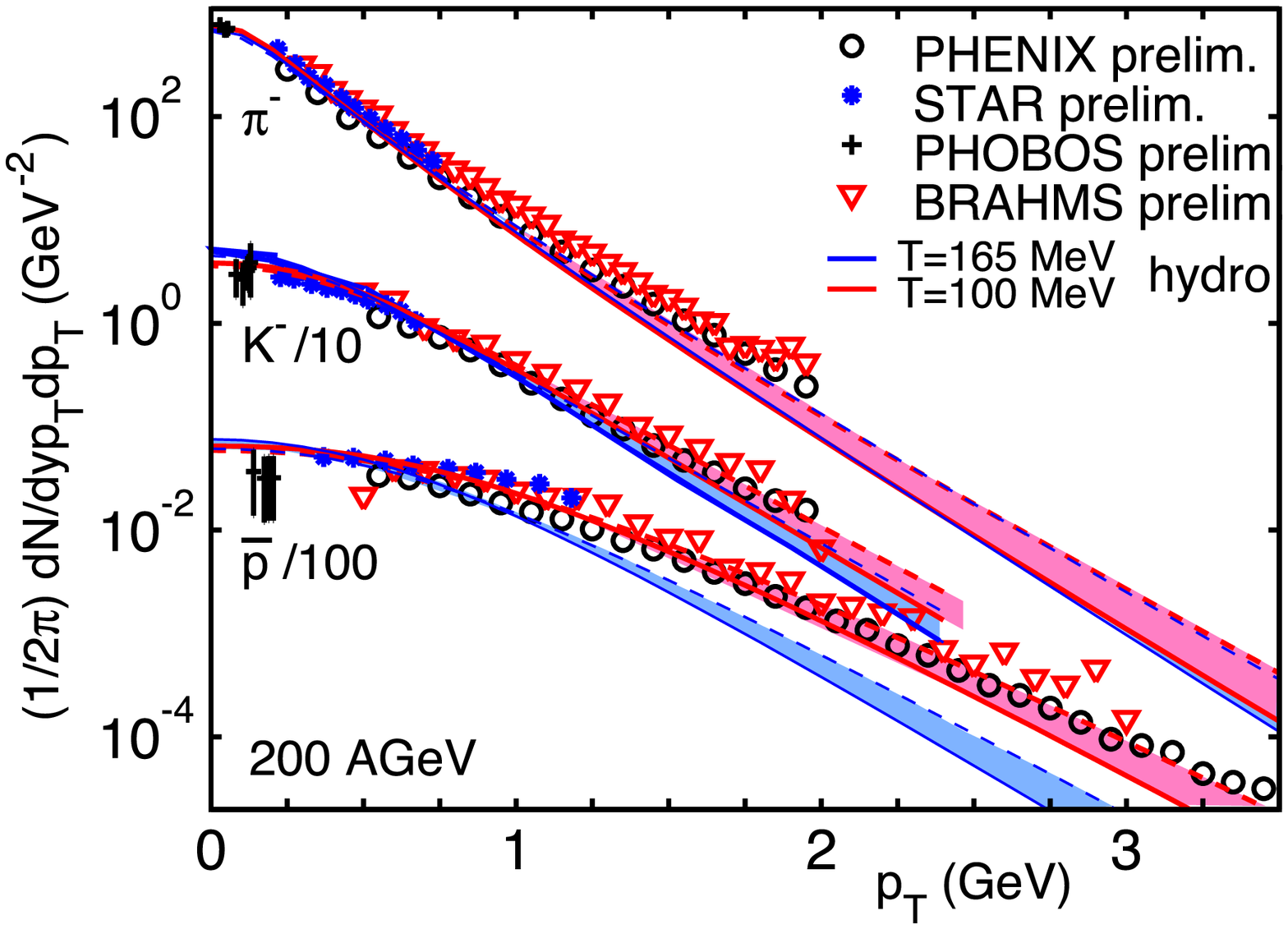}
\end{minipage}
\hspace*{3mm}
\begin{minipage}[t]{70mm}
\includegraphics[bb=8 30 513 405,width=70mm,height=51mm]{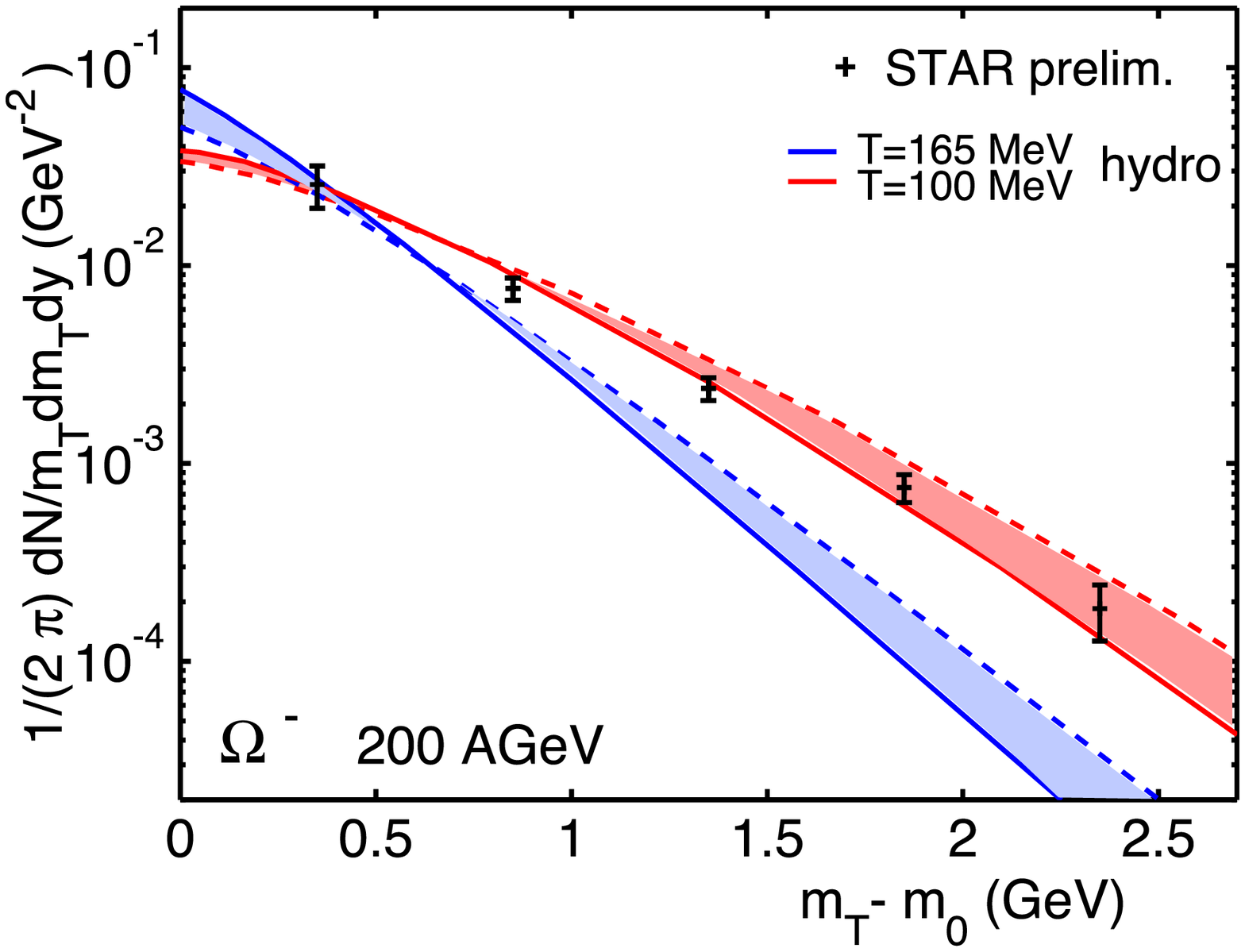}
\end{minipage}
\caption{\label{F1} 
(color online) Negative pion, kaon, antiproton, and $\Omega$ spectra from 
central Au+Au collisions at $\sqrt{s}\eq200\,A$\,GeV, as measured by the
four RHIC experiments \cite{PHENIX03spec200,STAR03spec200,%
PHOBOS03spec200,BRAHMS03spec200,STAR03omega}. The curves show hydrodynamical 
calculations as described in the text.}
\end{figure}
%%%%%%%%%%%%%%%%%%%%%%%%%%%%%%%%%%%%%%%%%%%%%%%%%%%%%%%%%%%%%%%%%%%%%%%%%%%%%%%

In Fig.~\ref{F1} I show the (absolutely normalized) single particle 
$p_\perp$-spectra for negatively charged pions, kaons and antiprotons
(left panel) as well as $\Omega$ baryons (right panel) measured in Au+Au 
collisions at RHIC together with hydrodynamical results \cite{KR03}. 
In order to illustrate the effect of additional radial flow generated by 
elastic scattering in the late hadronic stage below $T_{\rm cr}$, two sets 
of curves are shown: the lower (blue) bands correspond to kinetic decoupling 
at $T_{\rm cr}\eq165$\,MeV, whereas the upper (red) bands assume decoupling 
at $T_{\rm dec}\eq100$\,MeV. The width of the bands indicates the 
sensitivity of the calculated spectra to an initial transverse flow of
the fireball already at the time of thermalization: The lower end assumes 
no initial transverse flow whereas the upper end implements an initial 
radial flow profile $v_r(r_\perp,\tau_{\rm eq})\eq\tanh(\alpha r_\perp)$
with $\alpha\eq0.02$\,fm$^{-1}$ at $\tau_{\rm eq}\eq0.6$\,fm/$c$ (which
seems to be slightly preferred by the data). The hydrodynamic model 
output shows \cite{KSH00} that it takes about 9-10\,fm/$c$ until the
fireball has become sufficiently dilute to completely convert to hadronic
matter and another 7-8\,fm/$c$ to completely decouple. Figure~\ref{F1} 
shows clearly that by the time of hadronization hydrodynamics has not yet
generated enough radial flow to reproduce the measured proton and $\Omega$
spectra; these heavy hadrons, which are particularly sensitive to radial
flow effects, require the additional collective ``push'' created by
resonant (quasi)elastic interactions during the fairly long-lived hadronic 
rescattering stage between $T_{\rm cr}$ and $T_{\rm dec}$. Even though
independent flow fits to $\pi,K,p$ spectra on the one hand and 
multistrange hyperon spectra on the other \cite{Adams:2003fy} suggest
(within some systematic uncertainty related to the strong anticorrelation
between the extracted flow and temperature values) that $\Xi$ and $\Omega$ 
hyperons decouple slightly earlier (at somewhat higher temperature and 
with less radial flow) than pions, kaons, and protons, their 
{\em immediate} decoupling directly at hadronization is obviously not 
dynamically consistent with the successful hydrodynamic approach.

The strong flattening of the (anti)proton spectra by radial flow provides
a natural explanation for the (initially puzzling) experimental observation
that for $p_\perp{\,>\,}2$\,GeV/$c$ antiprotons become more abundant than
pions. For a hydrodynamically expanding thermalized fireball, at 
relativistic transverse momenta $p_\perp{\,\gg\,}m_0$ all hadron spectra 
have the same slope \cite{Lee:1990sk}, and at fixed $m_\perp{\,\gg\,}m_0$ 
their relative normalization is given by $(g_i\lambda_i)/(g_j\lambda_j)$ 
(where $g_{i,j}$ is the spin-isospin degeneracy factor and 
$\lambda_{i,j}=e^{\mu_{i,j}/T}$ is the fugacity of hadron species $i,j$). 
Due to the large antiproton chemical potential at kinetic freeze-out
which is required to maintain the $\bar p$ abundance at temperatures
below chemical freeze-out, this asymptotic $\bar p/\pi^-$ ratio is
predicted to be about 28 \cite{Kolb:2003dz}. Within the hydrodynamic
approach the surprising fact is therefore {\em not} that $\bar p/\pi{\,>\,}1$
at $p_\perp{\,>\,}2$\,GeV/$c$, but that this ratio seems to saturate
around 1 and never much exceeds this value \cite{PHENIX03spec200}.
As we will see below, this is a signature of the beginning breakdown
of the hydrodynamic model which stops working for baryons with
transverse momenta above about 2.5\,GeV/$c$.

As shown elsewhere (see Fig.~1 in \cite{Heinz:2002un}), once the hydrodynamic
model parameters have been fixed to describe pions, protons, and total
multiplicity in {\em central} Au+Au collisions, the model describes these 
and all other hadron spectra not only in central, but also in {\it peripheral}
collisions, up to impact parameters of about 10\,fm and with similar quality.
No additional parameters enter at non-zero impact parameter -- only
the initial conditions change due to the changing overlap geometry, but
this is completely accounted for by the Glauber model. Therefore, 
hydrodynamic results for the elliptic flow, discussed in the next Section,
are parameter-free predictions of the model.

%%%%%%%%%%%%%%%%%%%%%%%%%%%%%%%%%%%%%%%%%%%%%%%%%%%%%%%%%%%%%%%%%%%%%%%%%%%
\subsection{Hydrodynamic elliptic flow and RHIC data}
%%%%%%%%%%%%%%%%%%%%%%%%%%%%%%%%%%%%%%%%%%%%%%%%%%%%%%%%%%%%%%%%%%%%%%%%%%%

\begin{figure}[htb]
\hspace*{2mm}
\begin{minipage}[t]{70mm}
\includegraphics*[bb=61 210 568 594,width=70mm]{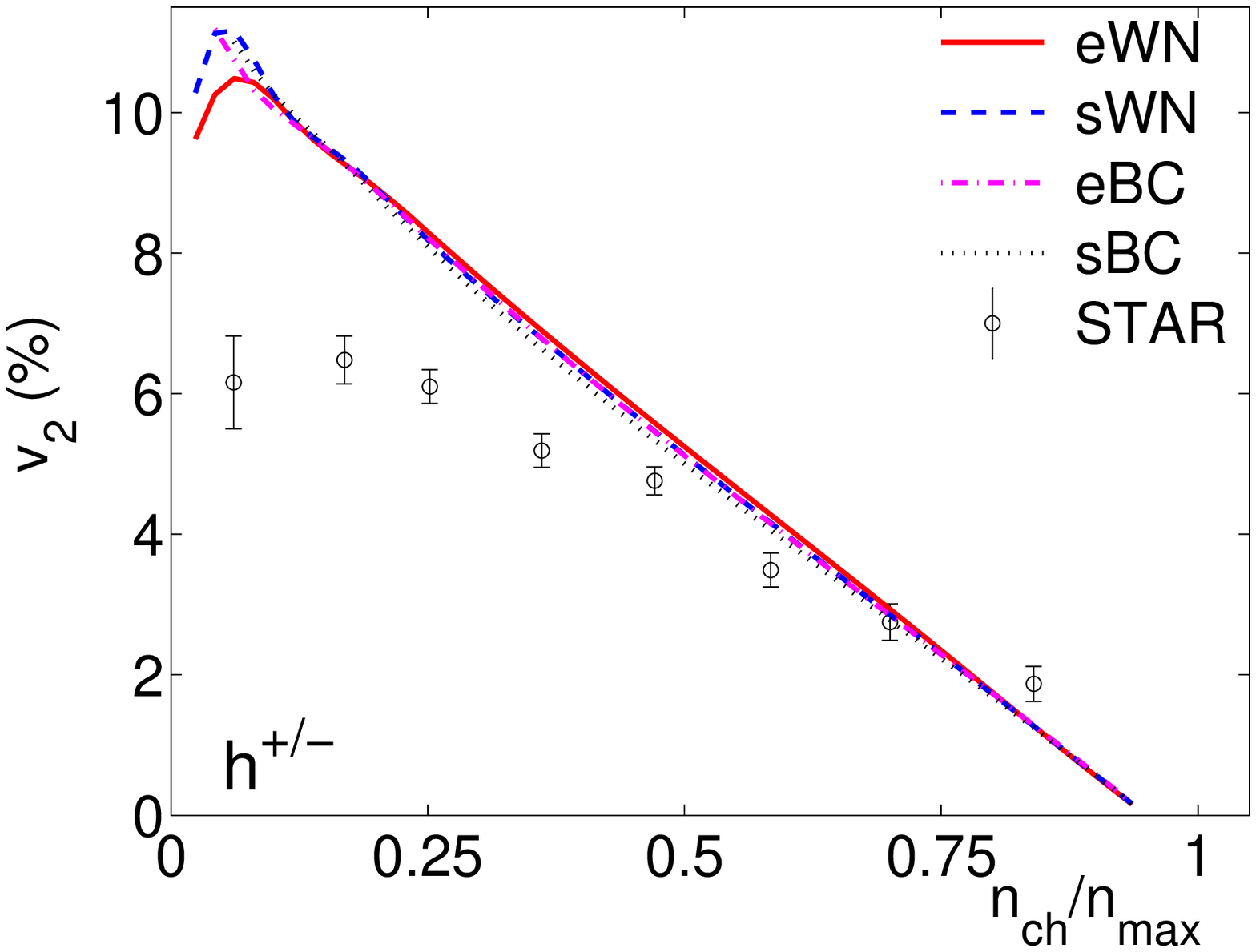}
\end{minipage}
\hspace*{1mm}
\begin{minipage}[t]{80mm}
\includegraphics*[bb=0 17 567 470,width=80mm,height=56.5mm]{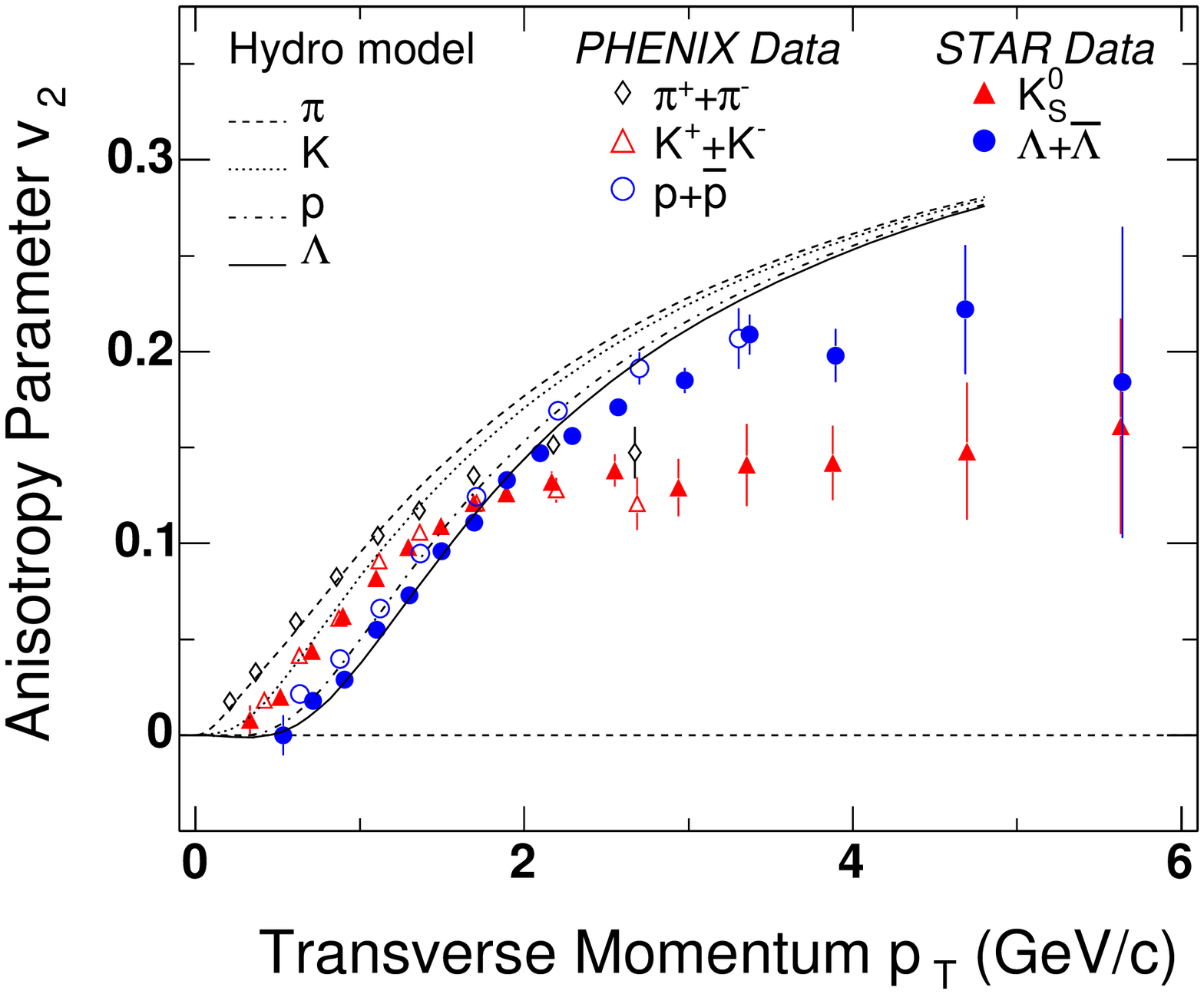}
\end{minipage}
\caption{\label{F2} 
(color online) Elliptic flow coefficient $v_2$ for all charged particles 
from 130\,$A$\,GeV Au+Au collisions (left panel \cite{KHHET}) and for 
several identified hadron species from 200\,$A$\,GeV Au+Au collisions 
(right panel \cite{Sorensen:2003kp}), compared with hydrodynamic predictions. 
The left panel shows the $p_\perp$-averaged elliptic flow as a function 
of collision centrality, parametrized by the 
charged multiplicity density $n_{\rm ch}$ at midrapidity ($n_{\rm max}$ 
corresponds to the largest value in central collisions). The right panel shows 
the differential elliptic flow $v_2(p_\perp)$ for minimum bias collisions. 
The data were collected by STAR and PHENIX  
\cite{Ackermann:2001tr,PHENIXv2,Adams:2003am}. The curves in the 
left panel are hydrodynamic calculations corresponding to different 
choices for the initial energy density profile (see \cite{KHHET} for 
details). The curves in the right panel were first published in 
\cite{Huovinen:2001cy}.
\vspace*{-3mm}
}
\end{figure}

Figure~\ref{F2} shows the predictions for the elliptic flow coefficient 
$v_2$ from Au+Au collisions at RHIC, together with the data 
\cite{Ackermann:2001tr,PHENIXv2,Adams:2003am}. For impact parameters 
$b{\,\leq\,}7$\,fm (corresponding to $n_{\rm ch}/n_{\rm max}{\,\geq\,}0.5$)
and transverse momenta $p_\perp{\,\leq\,}1.5{\,-\,}2$\,GeV/$c$ the data
are seen to exhaust the upper limit for $v_2$ obtained from the hydrodynamic
calculations. For larger impact parameters $b{\,>\,}7$\,fm the 
$p_\perp$-averaged elliptic flow $v_2$ increasingly lags behind the 
hydrodynamic prediction. It is tempting to attribute this to a lack of 
early thermalization when the initial overlap region becomes too small
\cite{Heinz:2001xi}. However, the work by Teaney \cite{Teaney:2001cw}, who 
coupled hydrodynamic evolution of the quark-gluon plasma above $T_{\rm cr}$
with a microscopic kinetic evolution of a hadronic resonance gas using
RQMD below $T_{\rm cr}$, showed that at least a strong contributing factor
%
%%%%%%%%%%%%%%%%%%%%%%%%% Fig. 2 %%%%%%%%%%%%%%%%%%%%%%%%%%%%%%%%%%%%%%%%%%%%
%%%%%%%%%%%%%%%%%%%%%%%%%%%%%%%%%%%%%%%%%%%%%%%%%%%%%%%%%%%%%%%%%%%%%%%%%%%%%%%
%
to the lack of elliptic flow in peripheral collisions (as well as in 
central collisions at lower energy) is the large viscosity {\em in the 
late hadronic stage}. If the initial energy density is not high
enough, the initial spatial deformation does not completely disappear
until hadronization. Whereas ideal hydrodynamics then predicts a continued
growth of the elliptic flow during the hadronic stage, driven by the still
existing anisotropies in the pressure gradients, realistic hadron cascades
produce very little additional elliptic flow, indicating that the hadronic
matter is very viscous. To completely saturate the elliptic flow value
predicted by {\em ideal} hydrodynamics thus requires not only very rapid
thermalization at the beginning of the reaction, but also a sufficiently
long {\em lifetime} of the thermalized QGP stage such that the initial 
spatial anisotropy has time to fully decay and the momentum anisotropy 
can fully develop before the system enters the viscous hadron gas phase.
This explains why the elliptic flow lags behind the hydrodynamic prediction
at RHIC in peripheral Au+Au collisions at midrapidity, in minimum bias 
Au+Au collisions at forward and backward rapidities  
\cite{Hirano02,Hirano:2001eu,Heinz:2004et}, as well as quite generally 
at lower collision energies \cite{NA49v2PRC,Agakichiev:2003gg}. From 
an analysis of all the available data we recently concluded 
\cite{Heinz:2004et} that an average initial energy density 
$\langle e\rangle_{\tau_{\rm eq}}{\,\geq\,}10$\,GeV/fm$^3$ appears
to be necessary to provide sufficient evolution time in the plasma phase
for full development of the momentum anisotropy before hadronization and 
therefore full saturation of the hydrodynamic elliptic flow limit.  
	
The $p_\perp$-dependence of $v_2$ (right panel of Fig.~\ref{F2}) tells 
another story: whereas the hydrodynamic model predicts a continuous rise
of $v_2$ with increasing $p_\perp$, the measured elliptic flow appears 
to saturate at high $p_\perp$. Baryon elliptic flow (here shown are 
protons and $\Lambda$ hyperons, but $\Xi$ and $\Omega$ hyperons follow 
the same systematics \cite{Castillo:2004jy}) saturates at a higher
value than meson (pion and kaon) elliptic flow. These saturation patterns
imply a break of the data away from the hydrodynamic prediction, and
again this break happens at higher $p_\perp$ for baryons 
($p_\perp^{\rm break}{\,\simeq\,}2.3-2.5$\,GeV/$c$) than for mesons 
($p_\perp^{\rm break}{\,\simeq\,}1.5$\,GeV/$c$). Below these break points, 
the elliptic flow of all hadrons measured so far is very well described
by hydrodynamics. In particular the hydrodynamically predicted {\em mass
splitting} of $v_2$ at low $p_\perp$ is perfectly reproduced by the data. 
Hydrodynamics thus gives an excellent description of {\em all} hadron
spectra below $p_\perp\eq1.5$\,GeV/$c$. Due to the exponentially falling
$p_\perp$ spectra, this includes more than 99\% of all produced hadrons,
so that it is fair to say that {\em the bulk of the fireball matter formed
in Au+Au collisions at RHIC behaves like a perfect fluid.}

In \cite{Huovinen:2001cy} we showed that the slope of $v_2(p_\perp)$ 
and its mass splitting are sensitive to the equation of state. The mass
splitting is larger for an EOS which includes a quark-hadron phase 
transition than for a pure resonance gas EOS extrapolated to arbitrarily 
high temperatures. Although requiring many more systematic model studies 
than so far available, these features offer the perspective of extracting 
detailed knowledge on the EOS during the early expansion stages from 
combined radial and elliptic flow measurements. So far we only know 
qualitatively (see Fig.~2 in \cite{Heinz:2002un} and the discussion 
in \cite{Teaney:2001cw}) that, for given angle-averaged $p_\perp$ 
spectra, the measured elliptic flow prefers an EOS that includes a 
phase transition at $T_{\rm cr}\eq165$\,MeV over one without such a 
transition. 

The excellent agreement between data and hydrodynamic model at transverse
momenta below about 2\,GeV becomes even more impressive after one begins 
to realize how easily it is destroyed: Parton cascade simulations with 
standard HIJING input generate almost no elliptic flow and require an 
artificial increase of the opacity of the partonic matter by a factor 
80 to reproduce the RHIC data \cite{Molnar:2001ux}. This shows that the 
QGP is much more strongly coupled than achievable within a standard 
perturbative QCD approach. Hadronic cascades of the RQMD and URQMD type 
(in which the high-density initial state is parametrized by 
non-interacting, pressureless QCD strings) predict \cite{Bleicher:2000sx} 
too little elliptic flow and a decrease of $v_2$ from SPS to RHIC, 
contrary to the data. Generically, models that don't include very 
strong rescattering in the very early fireball expansion stage fail 
to reproduce the elliptic flow data. 

As I will show in the next section, any significant delay of thermalization 
immediately leads to a significant loss of elliptic flow signal -- a fact 
that can be used to establish tight upper limits on the thermalization 
time scale. Even partial thermalization doesn't work: In 
\cite{Heinz:2002rs} we studied the transverse hydrodynamic evolution of 
a toy model in which only transverse momenta were thermalized while in 
longitudinal direction the fireball evolved ballistically. (One can think 
of this as implementing into the transverse hydrodynamics a very large 
value for the shear viscosity.) Since in this model there is no longitudinal
pressure, more work is done by the pressure in the transverse direction,
and one has a stiffer effective equation of state $p\eq{e}/2$ (instead 
of the ideal quark-gluon gas EOS $p\eq{e}/3$) driving the transverse
expansion. To obtain the same $p_\perp$-spectra as before one must thus
retune the initial conditions and let the hydrodynamic expansion start 
later, in order to avoid getting too much radial flow. The consequence 
of this is a reduction of the elliptic flow $v_2$ by almost a factor 2,
to levels far below the experimental data. This exercise demonstrates
a strong sensitivity of the elliptic flow to {\em early} shear viscosity. 
A complementary analysis by Teaney, who discussed {\em late-stage} viscous 
effects on hadron spectra, elliptic flow and HBT radii in \cite{T03},
arrives at similar conclusions. This sensitivity of $v_2$ to shear 
viscosity opens the perspective of limiting the shear viscosity of a 
quark-gluon plasma {\em experimentally} by performing systematic 
comparisons between elliptic flow data and {\em viscous} (i.e. non-ideal) 
hydrodynamics. This has recently led to renewed interest in practical 
implementations of viscous relativistic fluid dynamics \cite{Muronga}.

%%%%%%%%%%%%%%%%%%%%%%%%%%%%%%%%%%%%%%%%%%%%%%%%%%%%%%%%%%%%%%%%%%%%%%%%%%%%%
\subsection{Upper limits for the QGP thermalization time}
%%%%%%%%%%%%%%%%%%%%%%%%%%%%%%%%%%%%%%%%%%%%%%%%%%%%%%%%%%%%%%%%%%%%%%%%%%%%%

\begin{figure}[htb]
\includegraphics*[width=100mm]{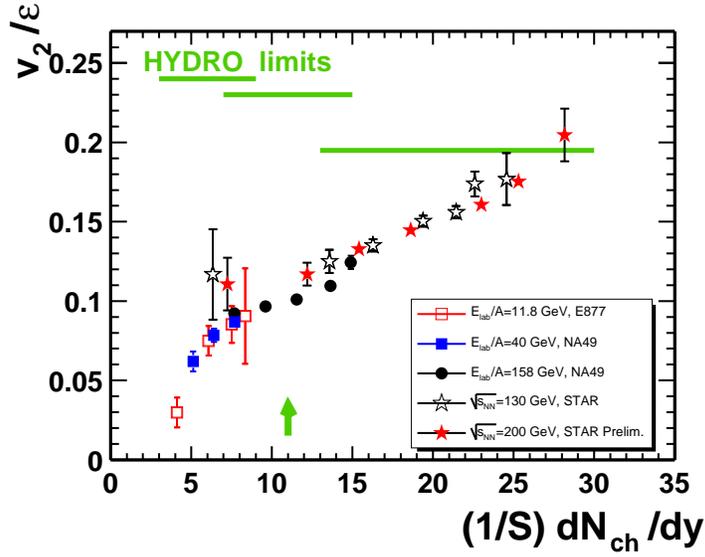}
\caption{\label{F3} 
(color online) Scaled elliptic flow $v_2/\epsilon_x$ as a function of 
$(1/S)\, dN_{\rm ch}/dy$ (i.e. the charged multiplicity density per
unit transverse overlap area $S$), for Pb+Pb and Au+Au collisions 
of different centralities at various collision energies, as compiled
by the NA49 Collaboration \cite{NA49v2PRC}. The lines indicate the
hydrodynamically predicted values at RHIC, SPS and AGS energies
\cite{KSH00} (see text for details).}
\end{figure}

The agreement between the elliptic flow data at RHIC and the hydrodynamic 
simulations can be used to set an upper limit on the thermalization time
scale in the dense partonic matter formed in the collision which is 
independent from the value $\tau_{\rm eq}\leq0.6$\,fm/$c$ obtained 
two chapters earlier by fitting the model to central collision spectra. 
Let me illustrate this by referring to Fig.~\ref{F3} which, to some extent,
is a different way of plotting the results shown in the left panel
of Fig.~\ref{F2}. In ideal hydrodynamics, the final elliptic flow $v_2$
%
%%%%%%%%%%%%%%%%%%%%%%%%% Fig. 3 %%%%%%%%%%%%%%%%%%%%%%%%%%%%%%%%%%%%%%%%%%%%
%%%%%%%%%%%%%%%%%%%%%%%%%%%%%%%%%%%%%%%%%%%%%%%%%%%%%%%%%%%%%%%%%%%%%%%%%%%%%%%
%
is directly proportional to the initial spatial deformation $\epsilon_x$,
with the proportionality constant being controlled by the equation of state
through the speed of sound $c_s^2\eq\partial p/\partial e$ \cite{O92}. 
The reason for this scaling is that the hydrodynamic equations of motion
are scale-free, and that therefore the initial condition for $\epsilon_x$
sets the scale for the final momentum anisotropy. This scaling assumes
that the momentum anisotropy can fully develop, i.e. that the hydrodynamic
evolution is not cut short by decoupling before the spatial anisotropy
$\epsilon_x$ has completely disappeared. As discussed in the previous 
chapter, this assumption is justified near midrapidity in central and 
near-central Au+Au collisions at RHIC, but breaks down more and more 
severely as one goes to more peripheral collisions, away from midrapidity, 
or to lower collision energies. This is at least part of the explanation
of the difference between the data points in Fig.~\ref{F3} and the 
horizontal lines labelled ``HYDRO limits'' \cite{Teaney:2001cw}.

These horizontal lines represent the fully developed hydrodynamic 
elliptic flow (scaled by the initial deformation $\epsilon_x$) for 
minimum bias Au+Au or Pb+Pb collisions for three different effective
sound speeds, reflecting the effective stiffness of our hadron gas 
+ quark-gluon plasma EOS for collisions at (from right to left or bottom 
to top) RHIC, SPS and AGS energies. They were extracted from the elliptic
flow excitation function for $b\eq7$\,fm Pb+Pb collisions given in 
Fig.~14 of \cite{KSH00}, noting that $b\eq7$\,fm is roughly equal to 
the average impact parameter in minimum bias Au+Au collisions. The 
RHIC value for $v_2/\epsilon_x$ is lower than the ones for SPS and
AGS energies since RHIC collisions spend more of the relevant time 
when the elliptic flow builds up in the soft phase transition region,
compared to SPS and AGS where most or all of that time is spent in
the stiffer hadron gas phase \cite{KSH00}. The increasing discrepancy
between these lines and the data as one decreases the collision energy
indicates the inability of the hadron gas phase (which gains more and
more dynamical weight) to respond to the spatial deformation of the
reaction zone. This is a late-stage effect and does not necessarily
indicate a lack of {\em early} thermalization.

However, some lack of early thermalization may still contribute to
this discrepancy. Let us, for the sake of the argument, concentrate on 
Au+Au collisions at $\sqrt{s}\eq200\,A$\,GeV (solid red stars in 
Figure~\ref{F3}) and pretend that, in fact, the {\em entire} discrepancy 
between data and theory is due to inefficient thermalization at the 
beginning of the collision. Let us model this inefficiency in a somewhat
black-and-white fashion, by assuming that the system evolves by 
free-streaming (i.e. without any interactions at all) for a time 
$\Delta\tau$, after which it instantaneously thermalizes and continues 
to evolve according to the laws of ideal fluid dynamics. As shown in
\cite{KSH00} (for a slightly more general argument see \cite{Kolb:2003dz}),
free-streaming results in a radial growth of the source which reduces 
its initial spatial deformation $\epsilon_x$ according to
\begin{equation} 
\label{epsilon}
  \frac{\epsilon_x(\tau_0{+}\Delta\tau)}{\epsilon_x(\tau_0)}
  = \left[ 
  1+\frac{(c\,\Delta\tau)^2}{\langle\bm{r}_\perp^2\rangle_{\tau_0}} 
    \right]^{\!-1}\,.
\end{equation}
Due to the hydrodynamic scaling of the final $v_2$ with the initial 
spatial deformation $\epsilon_x$, the final elliptic flow will be 
reduced by at least the same factor (and by even more if, after this 
delay, the hydrodynamic stage is no longer sufficiently long-lived to 
fully develop the elliptic flow). By assigning all of the ``missing'' 
elliptic flow in Figure~\ref{F3} to this ``delayed thermalization'' 
effect, we can therefore extract a conservative upper limit for the 
thermalization time scale $\Delta\tau$.

Note that even if (contrary to the above arguments based on Teaney's 
work \cite{Teaney:2001cw}) our assumption were true that {\em all} of 
the ``missing'' elliptic flow in Figure~\ref{F3} were due to inefficient
thermalization in the {\em initial partonic} stage (rather than in the 
{\em late hadronic} stage), this procedure will still only yield an upper 
limit for the thermalization time $\Delta\tau$: In real life, 
thermalization doesn't happen suddenly after a period of no scattering
at all, but gradually. Any type of scattering, however, will generate
anisotropic evolution (albeit not as strongly anisotropic as in ideal
fluid dynamics), i.e. faster expansion into the reaction plane than 
perpendicular to it, thereby reducing the spatial deformation $\epsilon_x$
(and thus the final $v_2$) {\em faster} than given by Eq.~(\ref{epsilon}). 

Let us now use this to extract an upper limit for the thermalization
time $\Delta\tau$ in $b\approx7$\,fm Au+Au collisions (4th solid red star
from the right in Fig.~\ref{F3}). According to the Figure, for this 
impact parameter $v_2/\epsilon_x$ is about 75\% of the hydrodynamic
value. Equating this with $\epsilon_x(\tau_0{+}\Delta\tau)/\epsilon_x(\tau_0)$,
and inserting $\langle r_\perp^2\rangle_{\tau_0}^{1/2}\eq3.5$\,fm for
$b\approx7$\,fm Au+Au, we find $\Delta\tau{\,\approx\,}2$\,fm/$c$ for 
this upper limit. For second solid red star from the right, the 
experimental value corresponds to 90\% of the hydrodynamic limit, and 
the resulting upper limit for the thermalization time is 
$\Delta\tau{\,\approx\,}1$\,fm/$c$. The right-most point in Figure~\ref{F3}
allows for a non-zero thermalization time only by virtue of its large
error bar (and, of course, systematic uncertainties in the exact
vertical position of the theoretical horizontal line).

We see that, even if we assign all of the ``missing'' elliptic flow 
to inefficient early thermalization, the resulting window for the
thermalization time scale is narrow and not larger than about 2\,fm/$c$
in minimum bias Au+A collisions and certainly less than 1\,fm/$c$ in
central Au+Au collisions. This window becomes even narrower when 
accounting for the well-documented \cite{Teaney:2001cw} inefficiency 
of the late hadronic stage to generate elliptic flow. When taken together,
these arguments yield very tight limits on the thermalization time scale
in the dense matter formed early in the collision which, I think,
are fairly represented (and not underestimated) by the value 
$\tau_{\rm eq}\eq0.6$\,fm/$c$ used as starting time in our hydrodynamic 
simulation. (Let me remind you once again that this time was determined
in a different way, using only information from central collisions, and
that it therefore enters my network of arguments for early thermalization
as an independent ``data point''.)

It is, of course, unsatisfactory to see in Figure~\ref{F3} the ideal
hydrodynamic limit just being reached at {\em one} point, namely in 
the {\em most} central Au+Au collisions at the {\em highest} available 
collision energy. The natural reaction of any serious physicist 
should be that this single point coincidence must be accidental, and 
that the Figure really shows incompatible tendencies in theory and 
data. The only problem with this ``conservative'' reaction is that it 
should be impossible to create more elliptic flow than predicted by 
ideal fluid dynamics! If at larger values for $(1/S)\,dN_{\rm ch}/dy$
the data for $v_2/\epsilon_x$ indeed continued to grow significantly
above the hydrodynamic limit (as suggested by the tendency of the data
in Figure~\ref{F3}), this would create {\em very serious} problems for 
theory. It is therefore very important to try to extend the range
of $(1/S)\,dN_{\rm ch}/dy$ to larger values, by studying either 
semi-central Pb+Pb collisions at the LHC or central U+U collisions
at RHIC \cite{KSH00,Heinz:2004et}. In order to support our present 
understanding of heavy-ion collision dynamics it will be essential to 
confirm that for $(1/S)\,dN_{\rm ch}/dy{\,>\,}30$ the ratio 
$v_2/\epsilon_x$ saturates at the hydrodynamic limit, showing only a 
very weak growth between RHIC and LHC (resulting from a stiffening of 
the effective equation of state) \cite{KSH00}.

%%%%%%%%%%%%%%%%%%%%%%%%%%%%%%%%%%%%%%%%%%%%%%%%%%%%%%%%%%%%%%%%%%%%%%%%%%%%%
\subsection{Hadronization via quark coalescence -- evidence for deconfinement}
%%%%%%%%%%%%%%%%%%%%%%%%%%%%%%%%%%%%%%%%%%%%%%%%%%%%%%%%%%%%%%%%%%%%%%%%%%%%%

We already mentioned that the right panel of Figure~\ref{F2} indicates
a characteristic difference between mesons and baryons in the way their
elliptic flow breaks away at intermediate $p_\perp$ from the hydrodynamic
behaviour at low $p_\perp$. In the same intermediate $p_\perp$ region a 
similar meson-baryon split is seen in the nuclear enhancement factors, 
specifically the ratio $R_{CP}$ between the various hadron yields (scaled 
by the number of binary nucleon-nucleon collisions) measured in central 
and peripheral Au+Au collisions \cite{Schweda:2004kd}. It is quite 
striking that in this ratio the $\phi$ meson data fall together with 
the much lighter kaons \cite{Schweda:2004kd} and not with the almost 
equally heavy protons \cite{Adler:2003kg} which instead follow the 
other measured baryons $\Lambda$, $\Xi$, and $\Omega$ 
\cite{Schweda:2004kd,Adler:2003kg}. This systematics indicates that 
hadron production at intermediate transverse momenta, 
$1.5\,{\rm GeV}/c{\,<\,}p_\perp{\,<\,}5-6$\,GeV/$c$, is controlled 
by the number of valence quarks inside the hadron rather than by its
mass. Mass scaling, as seen for the elliptic flow at low 
$p_\perp{\,\leq\,}1.5$\,GeV/$c$ (see Fig.~\ref{F2}, right panel), is 
characteristic 
of thermalization and hydrodynamic flow; a scaling with the number of 
valence quarks, on the other hand, is not consistent with hydrodynamics 
and rather suggests a quark coalescense picture \cite{Voloshin:2002wa}.
In fact, in the transition region around $1.5-2.5$\,GeV/$c$ where 
hydrodynamics begins to break down, a quark coalescence approach works 
extremely well, as I will now show. 

The coalescence model was previously developed for and applied to the
formation of deuterons and other light nuclei from nucleons in low-
and high-energy nuclear collisions (see \cite{Dover:1991zn,Scheibl:1998tk} 
and references therein). The formalism developed there is most useful if
the wave function of the cluster formed by the coalescence process
has a narrow internal momentum distribution (i.e. it is relatively
large). Clearly, hadrons have a much wider internal momentum distribution
than deuterons, but if the total hadron momentum is sufficiently large
(here we are interested in momenta ${>\,}1.5$\,GeV/$c$), neglecting
the internal relative momenta of the quarks inside the hadron may still
be a reasonable approximation. This is how the coalescence model has 
been applied to quark coalescence at RHIC by most authors 
\cite{Greco:2003xt,Fries:2003vb,Molnar:2003ff} (for exceptions see 
\cite{Lin:2002rw,Lin:2003jy}).

One starts from the following expressions for the invariant momentum 
spectrum for mesons $M$ and baryons $B$ \cite{Dover:1991zn,Scheibl:1998tk}:
\begin{eqnarray}
\label{coal}
  E\frac{dN_M}{d^3p} &=& \sum_{\alpha\bar\beta}
  \int \frac{p_\mu d\sigma^\mu(x)}{(2\pi)^3} 
  \int d^3q \, \left|\psi^{M,\alpha\beta}_{\bm{p}}(\bm{q})\right|^2 \,
  f_\alpha\left(\bm{p}+\half\bm{q},x\right)\, 
  f_{\bar\beta}\left(\bm{p}-\half\bm{q},x\right),
\\
\label{coal2}
  E\frac{dN_B}{d^3p} &=& \sum_{\alpha\beta\gamma}
  \int \frac{p_\mu d\sigma^\mu(x)}{(2\pi)^3} 
  \int d^3q_1 d^3q_2 
  \left|\psi^{B,\alpha\beta\gamma}_{\bm{p}}(\bm{q}_1,\bm{q}_2)\right|^2
\nonumber\\
  && \qquad\times\  
  f_\alpha\left(\bm{p}+\half\bm{q}_1,x\right)\, 
  f_\beta\left(\bm{p}+\half\bm{q}_2,x\right) \,
  f_\gamma\left(\bm{p}{-}\half(\bm{q}_1{+}\bm{q}_2),x\right).
\end{eqnarray}
The $\psi$'s are the internal wavefunctions of the hadrons, expressed 
as functions of the relative momenta of their valence quarks, and the
$f$'s are the valence quark phase-space distributions. The sums
extend over all spin, isospin and color channels contributing to the 
desired hadronic final state. The outer integral goes over the 
hadronization surface $\sigma(x)$. In the narrow wave function limit 
$\bm{q}_i\eq0$, and under certain assumptions about $\sigma(x)$ (e.g. 
that its spatial anisotropy is small so that the momentum anisotropy 
of the spectra is dominated by the momentum anisotropy of the quark 
distribution functions $f_{\alpha,\beta,\gamma}(\bm{p},x)$), these 
formulae imply that the azimuthal distributions of the hadrons are 
related to those of the quarks by 
\begin{equation}
   \frac{dN_M}{d\phi_p}\sim\left(\frac{dN_q}{d\phi_p}\right)^2,
     \quad
     \frac{dN_B}{d\phi_p}\sim\left(\frac{dN_q}{d\phi_p}\right)^3,
\end{equation}
and that the elliptic flow coefficients (neglecting small 
nonlinear terms) satisfy \cite{Molnar:2003ff}
\begin{eqnarray}
\label{v2}
  v_2^M(p_\perp) \approx 
  v_2^\alpha     \left(\textstyle{\frac{p_\perp}{2}}\right) + 
  v_2^{\bar\beta}\left(\textstyle{\frac{p_\perp}{2}}\right),\quad
  v_2^B(p_\perp) \approx
  v_2^\alpha\left(\textstyle{\frac{p_\perp}{3}}\right) + 
  v_2^\beta \left(\textstyle{\frac{p_\perp}{3}}\right) +
  v_2^\gamma\left(\textstyle{\frac{p_\perp}{3}}\right).
\end{eqnarray}
If all quark flavours carry the {\em same} elliptic flow, this simplifies 
further to \cite{Molnar:2003ff}
\begin{equation}
  v_2^h(p_\perp) \approx n\ v_2^q\left(\textstyle{\frac{p_\perp}{n}}\right).
\end{equation}
In other words, if we divide $v_2$ for a hadron by the number $n$ of 
its valence quarks and plot it against the scaled transverse momentum 
$p_\perp/n$, we should obtain a universal function 
$v_2^q\left(\textstyle{\frac{p_\perp}{n}}\right)$ describing the elliptic
flow of the quarks just before hadronization.

%%%%%%%%%%%%%%%%%%%%%%%%% Fig. 4 %%%%%%%%%%%%%%%%%%%%%%%%%%%%%%%%%%%%%%%%%%%%
\begin{figure}[htb]
\begin{minipage}[t]{60mm}
\includegraphics*[bb=20 20 575 584,width=60mm]{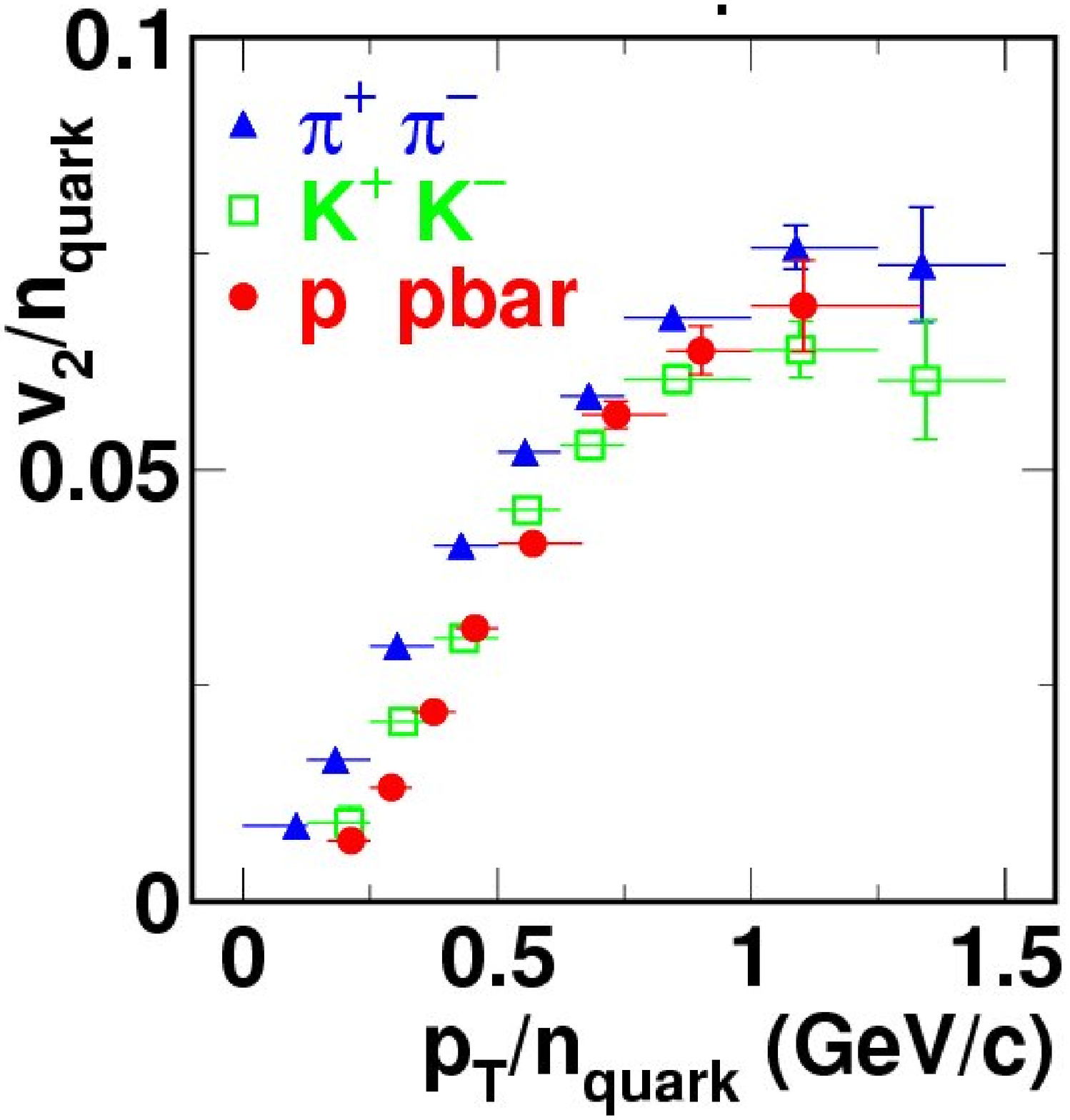}
\end{minipage}
%
%\hspace*{3mm}
\begin{minipage}[t]{90mm}
\includegraphics*[bb= 7 29 580 440,width=90mm,height=65mm]{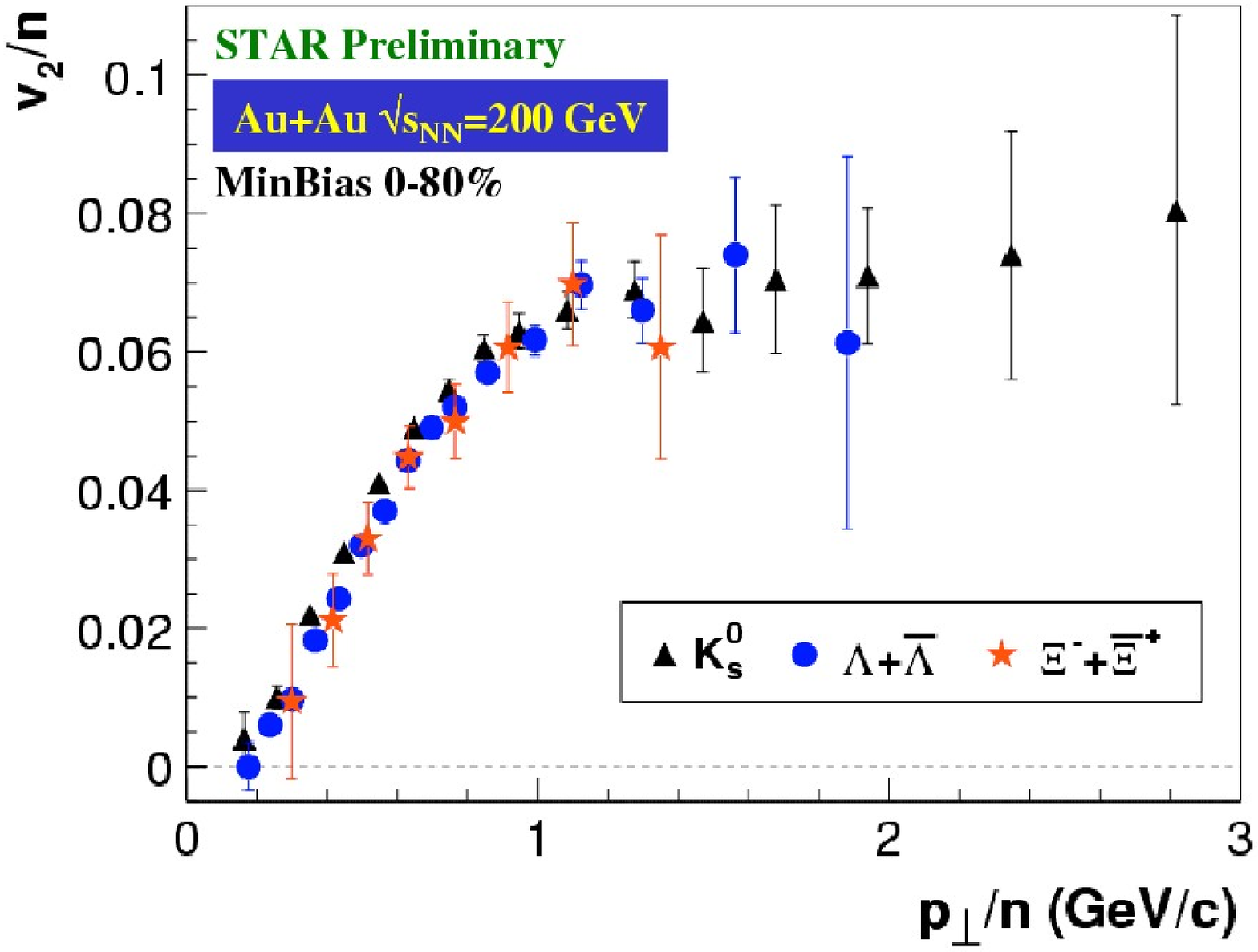}
\end{minipage}
\caption{\label{F4} 
(color online) Elliptic flow per valence quark, $v_2/n$, as a function
of transverse momentum per valence quark, $p_\perp/n$, for pions, kaons,
protons, $\Lambda$ and $\Xi$ hyperons and their antiparticles. Data are
from 200\,$A$\,GeV Au+Au collisions measured by PHENIX \cite{PHENIXv2} 
(left panel) and STAR \cite{Castillo:2004jy} (right panel). Universality
of this curve suggests that it represents the partonic elliptic flow
$v_2^{\rm parton}(p_\perp^{\rm parton})$, with no apparent difference 
between light and strange quark elliptic flow.}
\end{figure}
%%%%%%%%%%%%%%%%%%%%%%%%%%%%%%%%%%%%%%%%%%%%%%%%%%%%%%%%%%%%%%%%%%%%%%%%%%%%%%%
%

Figure~\ref{F4} shows that this universal scaling works excellently.
There is no indication for different elliptic flows for light and 
strange quarks. The common parton elliptic flow shows the same generic
form as the individual hadronic curves, following the hydrodynmic rise
at low $p_\perp$ up to about $p_\perp\eq750$\,MeV where it breaks away
from hydrodynamics, eventually saturating at a level of about 7\% (for 
minimum bias Au+Au collisions) above $p_\perp\eq1$\.GeV. Note that, on
the parton level, hydrodynamics works only out to about twice the 
mean transverse momentum; the coalescence mechanism ensures that this 
is sufficient for mesons and baryons to behave hydrodynamically to much
larger $p_\perp$, namely about 1.5\,GeV for mesons and about 2.3\,GeV
for baryons. Pentaquarks \cite{penta} should therefore show hydrodynamic
behaviour up to $p_\perp\eq3.8$\,GeV/$c$, and their elliptic flow
should saturate at 35\%!

Another important remark is that the extraction of the partonic flow 
from the rescaled hadronic elliptic flows is only possible because 
hydrodynamics breaks down at high $p_\perp$. If the elliptic flow 
continued to follow the almost linear rise to higher $p_\perp$, 
rescaling both vertical and horizontal axes would simply reproduce 
this linear curve, and nothing could be learned. Since breakdown of 
hydrodynamics produces a structure (shoulder) in the function 
$v_2(p_\perp)$ and this shoulder moves under rescaling of the axes, 
the model can be tested and verified. The strongest 
evidence for the coalescence model comes from the fact that the shoulder
of $v_2$ happens for different collision centralities at different 
absolute values for $v_2$ and at slightly different $p_\perp$ values;
however, the quark number scaling of the shoulder position works at
all impact parameters, i.e. different hadronic $v_2$-curves collapse
onto a single partonic $v_2$ curve at each collision centrality  
\cite{Sorensen:2003kp}.

Careful readers will note that for $p_\perp{\,<\,}750$\,MeV the 
extracted partonic $v_2$ is not quite universal. In particular the 
differential partonic elliptic flow extracted from pions seems to be 
higher than that from the other hadrons (left panel in Figure~\ref{F4}). 
But this is only a reflection of the hydrodynamic mass splitting 
(remember: hydro works at low $p_\perp$!) which, of course, cannot 
be scaled away by rescaling the axes. The coalescence model does {\em not} 
work at low $p_\perp$ (in particular, it violates entropy conservation 
because it reduces the number of entropy carriers by more than a factor 2!);
it can only be used at sufficiently high $p_\perp$ where coalescence
is a relatively rare process \cite{Molnar:2003ff}.

Clearly, the coalescence model makes essential use of the concept of 
deconfined (uncorrelated) quarks with independent phase-space 
distributions. I do not see how the fact that it works so perfectly 
can be interpreted in any other way than to provide at least 
indirect evidence for quark deconfinement in the early partonic phase.

%%%%%%%%%%%%%%%%%%%%%%%%%%%%%%%%%%%%%%%%%%%%%%%%%%%%%%%%%%%%%%%%%%%%%%%%%%%%%
\subsection{Conclusions: Have we seen the QGP?}
%%%%%%%%%%%%%%%%%%%%%%%%%%%%%%%%%%%%%%%%%%%%%%%%%%%%%%%%%%%%%%%%%%%%%%%%%%%%%

As the astute reader of these pages must have already concluded from my 
repeated use of the acronym QGP for the dense early collision stage, my
own answer to this question is: {\em YES!} The quantitative success of the 
hydrodynamic model in describing the bulk of particles emitted from 
Au+Au collisions at RHIC provides compelling
proof that thermalized matter at unprecedented energy densities 
$\langle e\rangle {\,>\,}100\,e_{\rm nm}$ has been created.
($e_{\rm nm}\eq0.125$\,GeV/fm$^3$ is the energy density of cold nuclear 
matter inside atomic nuclei.) There is only one known viable theoretical 
concept that can describe {\em thermalized} matter at such energy 
densities: {\em the Quark-Gluon Plasma}. The mass splitting of the elliptic
flow supports (but at the present accuracy does not prove) the hypothesis
that this plasma is separated from normal hadronic matter by a phase
transition (i.e. a soft region in the equation of state). The validity 
of the ideal fluid hydrodynamic model implies very severe constraints 
for the thermalization time scale and transport properties of this plasma: 
the required $\tau_{\rm eq}{\,<\,}1$\,fm/$c$ cannot be obtained with standard 
perturbative QCD scattering mechanisms and requires (presently unknown)
non-perturbative physics which makes the QGP created at RHIC behave like
a strongly interacting, almost ideal fluid rather than a weakly interacting
parton gas. 

One might worry that such non-perturbative processes might also modify 
the color deconfinement and chiral symmetry properties usually associated 
with the QGP, and transverse flow patterns do not, of course, provide 
any direct information about these characteristics. I showed that indirect 
arguments for color deconfinement can be extracted from the pattern in 
which the hydrodynamic description of hadron spectra begins to break down 
at intermediate transverse momenta: This pattern is consistently described 
by a quark-coalescence model, in which collective flow patterns are 
first imprinted on the momentum distributions of deconfined (uncorrelated)
quarks in the early partonic phase and then transferred to the late 
hadronic phase by a quark recombination mechanism. 

Another indirect argument going in 
the same direction is based on the success of the grand canonical 
thermodynamic ensemble in describing all the measured hadron abundance 
ratios \cite{BMMRS01}. This approach works extremely well even for 
multistrange hyperons and antihyperons, with a fully equilibrated 
strange particle sector (in contrast to elementary particle collisions
and nuclear collisions at lower energies where strangeness is found to
be suppressed relative to the grand canonical approach and requires 
taking into account, via canonical constraints and/or additional 
strangeness suppression factors, local correlations between strange 
quarks and antiquarks \cite{Braun-Munzinger:2003zd}). This implies 
that, say, in the production of an $\Omega$ hyperon, strangeness can 
be balanced by kaons or anti-hyperons {\em anywhere} in the 
fireball volume, including its far opposite edge. In other words,
any memory of the fact that the microscopic strangeness production
process in QCD is {\em local}, i.e. $s$ and $\bar s$ are created 
pairwise in a point, has been completely lost by the time strange 
hadrons form, and at hadronization the strange quarks and antiquarks 
are completely uncorrelated and equally likely to be found at any point 
in the fireball. It is very difficult, if not impossible, to explain 
this without invoking strong rescattering dynamics among deconfined
quarks and gluons before the onset of hadron formation.

%%%%%%%%%%%%%%%%%%%%%%%%%%%%%%%%%%%%%%%%%%%%%%%%
%% BACKMATTER
%%%%%%%%%%%%%%%%%%%%%%%%%%%%%%%%%%%%%%%%%%%%%%%%

%\begin{theacknowledgments}
\subsection{Acknowledgements}
The author thanks the organizers of the school and workshop for their
warm hospitality and for generating an intellectually very stimulating 
environment. This work was supported by the U.S. Department of Energy
under contract DE-FG02-01ER41190.
%\end{theacknowledgments}

%% If the bibliography is produced without BibTeX comment out the


%\bibliography{sample}

%  \typeout{******************************************}
%  \typeout{** Please run "bibtex \jobname" to optain}
%  \typeout{** the bibliography and then re-run LaTeX}
%  \typeout{** twice to fix the references!}
%  \typeout{******************************************}

%%%%%%%%%%%%%%%%%%%%% References %%%%%%%%%%%%%%%%%%%%%%%%%%%%%%%%%%%%%

\end{document}